\documentclass[pra,10pt,twocolumn,superscriptaddress,floatfix,showpacs]{revtex4-1}

\usepackage{subfig}
\usepackage{epstopdf}
\usepackage[utf8]{inputenc}
\usepackage[T1]{fontenc}     
\usepackage[british]{babel}  
\usepackage[sc,osf]{mathpazo}
\usepackage[scaled=0.86]{berasans}  
\usepackage[colorlinks=true, citecolor=blue, urlcolor=blue]{hyperref}  
\usepackage{graphicx} 
\usepackage[babel]{microtype}  
\usepackage{amsmath,amssymb,amsthm,bm,amsfonts,mathrsfs,bbm} 

\usepackage{xspace}  
\usepackage{pgfplots}
\usepackage{xcolor,colortbl}
\def\ba{\begin{equation}}
\def\ea{\end{equation}}
\def\bea{\begin{eqnarray}}
\def\eea{\end{eqnarray}}
\def\ben{\begin{equation*}}
\def\een{\end{equation*}}
\def\bean{\begin{eqnarray*}}
\def\eean{\end{eqnarray*}}
\def\bma{\begin{mathletters}}
\def\ema{\end{mathletters}}
\def\bi{\begin{itemize}}
\def\ei{\end{itemize}}

\newcommand{\be}{\begin{equation}}
\newcommand{\ee}{\end{equation}}

\newcommand{\kommentar}[1]{}

\newcommand{\forget}[1]{}

\begin{document}
\title{Nontrilocality: Exploiting nonlocality from three particle systems}
\author{Kaushiki Mukherjee}
\email{kaushiki_mukherjee@rediffmail.com}
\affiliation{Department of Mathematics, Government Girls’ General Degree College, Ekbalpore, Kolkata-700023, India.}

\author{Biswajit Paul}
\email{biswajitpaul4@gmail.com}
\affiliation{Department of Mathematics, South Malda College, Malda, West Bengal, India}

\author{Debasis  Sarkar}
\email{dsappmath@caluniv.ac.in, dsarkar1x@gmail.com}
\affiliation{Department of Applied Mathematics, University of Calcutta, 92, A.P.C. Road, Kolkata-700009, India.}


\begin{abstract}
In Phys. Rev. Lett. \textbf{104},170401 (2010), Branciard \textit{e.t al.} first characterized the correlations arising in an entanglement swapping network under the assumption that the sources generating the initially uncorrelated quantum systems are independent. Precisely speaking, in Phys. Rev. Lett. \textbf{104},170401 (2010) and later in Phys. Rev. A \textbf{85},032119 (2012) the authors analyzed the importance of \textit{bilocal}(source independence) assumption to lower down the restrictions over correlations for revealing quantumness in the network where each  of two sources generates a bipartite entangled state. In this context one may find interest to characterize correlations in a network involving independent sources which can correlate more than two initially uncorrelated multipartite entangled quantum systems. Our present topic of discussion basically analyzes such a network scenario. Specifically we introduce \textit{trilocal network scenario} where each of three sources  independently generates a tripartite entangled quantum system thereby exploring the role of source independence assumption to exploit nonlocality in a network involving multipartite entanglement analogous to bilocal assumption in a network where only bipartite entanglement was considered. Interestingly, genuine entanglement content did not turn out to be an essential requirement for exploiting nonlocality in such a scenario. Moreover it is interesting to explore whether such a scenario can be generalized so as to characterize correlations arising in a network involving $n$ number of $n$ partite systems for any finite value of $n\geq4$ under source independence assumption.
\end{abstract}

\date{\today}

\maketitle
\section{introduction}
The study of quantum nonlocality, mainly dealing with analysis of correlated statistics between outcomes of measurements on space-like separated parties, formally originated with the seminal work of J.S.Bell\cite{Bel,Bel2}. The basic assumption behind Bell's idea of local causality was that correlations between distant particles result from causal influences originating in their common past. However predictions from quantum theory can explain certain correlations arising due to local measurements on entangled particles which are inexplicable by any theory where outcomes arising due to local measurements by separated parties are determined by variables correlated at the source\cite{Brunreview}. Violation of Bell inequalities detects such nonlocal behavior of entangled states. But apart from this remarkable behavior of quantum measurements establishing nonlocal correlations, there exists yet another even more striking feature of quantum measurements: quantum theory predicts that suitable measurements can generate nonlocal correlations even from particles that never  interacted directly. Such nonclassical nature of quantum theory is revealed through a process called \textit{entanglement swapping} \cite{ES1}. \\
Entanglement swapping procedure basically involves three parties, say Alice, Bob and Charlie and two independent pairs of entangled particles, say $\rho_{AB}$ and $\rho_{AC}$ such that initially  starting from these two uncorrelated pairs of entangled particles, one of the three parties, say Bob can measure jointly one particle from each pair, so that the remaining two particles become entangled($\rho_{AC}$), even though they share no direct common past(however Bob's  communication of output of his joint measurement to Alice and Charlie acts as a common past for Alice and Charlie's particles). So now depending on the joint measurement performed by Bob, the resulting bipartite conditional state $\rho_{AC}$(conditioned on Bob's output), shared between Alice and Charlie, is an entangled pair and can exhibit nonlocality via violation of Bell inequalities. So, intuitively this procedure seems to exhibit nonlocal effects more strongly compared to standard Bell tests\cite{Cl}. Based on this intuition several research works have dealt with characterization of correlational statistics involved in an entanglement swapping procedure. In recent times, \textit{Bilocal Scenario}(see FIG.1), a  more general scenario of three parties characterized by source independence(\textit{bilocal} assumption) was introduced in \cite{BRA}. An entanglement swapping procedure forms a special case of this scenario. \\
 Bilocality assumption can be applied to lower down the requirements to demonstrate quantumness in a system compared to some pre-existing standard procedures. For instance, in \cite{BRAN} Rosset \textit{e.t al.} showed that in a standard entanglement swapping involving two independent pair of entangled Werner states\cite{Wer}, a visibility of \textit{V} $> 50\%$ is enough to reveal quantumness (nonbilocality)  whereas nonlocal correlations are generated in the network for a visibility of \textit{V}$> 70.7\%$(for projective measurements\cite{Wer}) in usual Bell sense. This in turn guaranteed presence of local but nonbilocal correlations in a quantum network. In course of time bilocal scenario has been modified and analyzed in various aspects\cite{Tav,km,raf,den,kmb}. In \cite{Tav} $n$- local scenario($n$ independent sources) was introduced where the authors considered a star network. In \cite{km}, $n-$ local scenario was dealt with in a linear chain. In \cite{raf}, the authors introduced some more generalized scenarios involving complex Bayesian networks(graphical model to explain probabilistic causal scenarios). \\
As has already been discussed before, the most striking feature of entanglement swapping procedure is its ability to generate nonclassical correlations between particles that never directly interacted together. So from that perspective it becomes interesting to explore networks where the number of non interacting particles(say $N$) can be increased. In the original bilocal scenario\cite{BRA}, the number of such non interacting particles was only $2$ as in any standard entanglement swapping network(see FIG.1). $N$ was increased from to $2$ to $m$(say) such that $m$$>$$2$ where the authors used star network configuration\cite{Tav}. In such a network, however each independent source(shared between the central party of the network and one edge party \cite{Tav}) generated a bipartite entangled state.  Due to a complex structure of multipartite entanglement\cite{review} in contrast to bipartite entanglement, it becomes more interesting to study nature of correlations generated in the network when multipartite entangled states can be used in this context which in turn is obviously helpful to develop a better understanding of the interplay between entanglement and nonlocality. In our present work we intend to contribute in this direction. Precisely speaking, we have introduced a network scenario consisting of three independent sources such that each source generates a tripartite system(see FIG.2) thereby studying nature of correlations and other related issues. Interestingly such a network can be generalized so that the network now involves $n$ independent sources, each generating an $n$ partite system(see FIG.6).  \\
Rest of our paper is organized as follows:  in Sec.\ref{mot} we discuss the motivation of our present work. In Sec.\ref{pre} we deal with some basic prerequisites. In Sec.\ref{tri} we introduce the \textit{$n$-local network scenario} for $n$$=$$3$, i.e., $\textit{trilocal network scenario}$ followed by derivation of the Bell-type inequalities. In Sec.\ref{qu}, we discuss the quantum violation obtained in the network. In Sec.\ref{res} we deal with the advantage of source independence assumption to exploit nonclassical feature of quantum correlations compared to some pre-existing methods of doing so. We then generalize the trilocal network scenario in Sec.\ref{network}. We finally conclude in Sec.\ref{dis} discussing our findings in a nutshell along with related topics for future research.
\section{motivation}\label{mot}
The increase in the number of non interacting particles($N$) together with use of multipartite entanglement in an entanglement swapping network basically motivates our present topic of discussion. To be specific, here we mainly focus on tripartite entangled states. However the discussion can be extended for $n-$ partite entangled states. For our purpose we first introduce a general network scenario involving five parties and three independent sources(see FIG.2). We define such a network, characterized by five partite correlation terms as a \textit{trilocal network scenario}. We design a set of Bell-type inequalities necessarily satisfied under source independence assumption known as $\textit{trilocal assumption}$. There exist families of tripartite entangled quantum states, both pure and mixed which can generate nontrilocal correlations and hence capable of exhibiting nonlocality(apart from standard Bell sense). Interestingly, genuine entanglement is not a necessary requirement for generation of nontrilocality in a quantum network. This form of nonclassical phenomenon can also be observed in a network using biseparable quantum states.  \\
As will be discussed in details later, a network scenario is mainly characterized by correlation terms.  In this context we have also designed a set of Bell-type inequalities such that local correlations generated in a network necessarily satisfy this set of inequalities. Hence violation of at least one inequality from the set detects any nonlocal behavior of the correlations generated in the network. Just as in the case of bilocal assumption, requirements to demonstrate quantumness in a network is reduced under trilocal assumption. To be specific, we observed that for some families of mixed entangled tripartite states, correlations satisfy the local inequality and hence may not be nonlocal but are however nontrilocal. Moreover nontrilocal correlations are observed in an entanglement swapping network(characterized by source independence), even if  after completion of  the swapping procedure and communication of outputs of swapping from the parties(say Bob and Charlie) performing joint measurements to the remaining parties(say Alice, Dick and Tom), local correlations are shared between Alice, Dick and Tom. To be precise, nonlocality can be guaranteed in an entanglement swapping network due to generation of nontrilocal correlations even if each of the swapped states shared between Alice, Dick and Tom is local in some specific Bell scenario\cite{Sliwa}. In this context, the most important utility of source independence assumption is to increase resistance to noise of noisy states encountered in practical tasks. Besides, we also encountered some instances where the states, resulting from joint measurements by each of two parties and then shared between remaining three parties are atmost ppt bound entangled yet nontrilocal correlations are generated in the network. This observation gives rise to somewhat counterintuitive feature of quantum correlations in a network. This is due to the fact that entangling never interacting particles via swapping procedure seems to be responsible for generation of any sort of nonclassical correlations in a quantum network. However such a feature of quantum correlations can be attributed to source independence assumption in the quantum network. All these observations, being related to an experimental scenario(entanglement swapping network and other quantum networks) are expected to be significant from practical perspectives and thereby are supposed to be contributory in the field of information processing tasks, like  Quantum Key Distribution(QKD)\cite{Acin,Mayer,key1,key2}, private randomness generation\cite{Pironio,Colbeck}, device independent entanglement witnesses\cite{Bancal}, Bayesian game theoretic applications \cite{game} etc. Besides, our network scenario, being related to entanglement swapping networks can be experimentally verified. \\
After explicitly discussing the motivations behind our work, we now move to present our scenario along with related observations. But before defining our premise we provide with some necessary preliminaries required to present our work. \\

\section{Preliminaries}\label{pre}
\subsection{Bilocal Scenario of Two-Qubit States}
Consider the bilocal experimental setup (FIG.1) as depicted in \cite{BRA,BRAN}. It is a network of three parties Alice($A$), Bob($B$) and Charlie($C$) and two sources $S_1$ and $S_2$ all arranged in a linear pattern such that a source is shared between any pair of adjacent parties. The two sources $S_1$ and $S_2$ are independent to each other($\textit{bilocal assumption}$). A physical system represented by $\lambda_1$ and $\lambda_2$ is send by $S_1$ and $S_2$ respectively. Independence of the two variables $\lambda_1$ and $\lambda_2$ is guaranteed by independence of $S_1$ and $S_2$. All parties can perform dichotomic measurements on their systems labeled by $x,\,y,\,z$ for Alice, Bob and Charlie and their outcomes are denoted by $a,\,b,\,c$ respectively. In particular, Bob may perform a joint measurement on the joint state of the two systems that he receives from $S_1$ and $S_2$. The correlations obtained in the network are local if they take the form:
$p(a, b, c|x, y, z)=\iint d\lambda_{1} d\lambda_{2} {\rho(\lambda_1,\lambda_2)}$
\begin{equation}\label{p11}
P(a|x, \lambda_1)P(b|y, \lambda_1, \lambda_2)P(c|z, \lambda_2)
\end{equation}
The tripartite correlations are bilocal if they have a decomposition in the above form(Eq.(\ref{p11})) together with the constraint:
\begin{equation}\label{p2}
    \rho(\lambda_1,\lambda_2)=\rho_1(\lambda_1)\rho_2(\lambda_2)
\end{equation}
 imposed on the probability distributions of the hidden states $\lambda_1, \lambda_2$. In particular, Eq.(\ref{p2}) refers to the \textit{bilocal constraint}.
\begin{figure}[htb]
\includegraphics[width=3in]{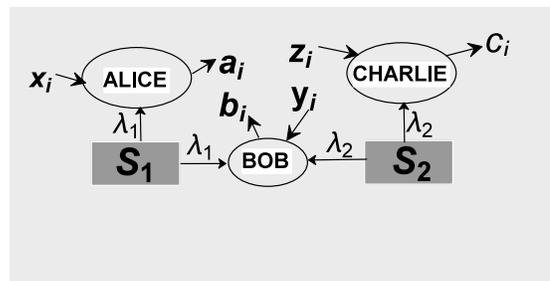}\\
\caption{\emph{Schematic diagram of bilocal scenario\cite{BRA,BRAN}. }}
\end{figure}
An entanglement swapping network is a particular case of this scenario\cite{BRA,BRAN,Ross,Fritz,Ros}.\\
 The correlations of the form (Eq.(\ref{p11})) and (Eq.(\ref{p2})) are bilocal if they satisfy the inequality\cite{BRAN}:
\begin{equation}\label{A3}
 \sqrt{|I|} + \sqrt{|J|}\leq1
\end{equation}
$\textmd{with}\,\,I$=$\frac{1}{4}\sum \limits_{x, z=0,1}\langle A_x B_0 C_z\rangle,\,J$=$\frac{1}{4}\sum \limits_{x, z=0, 1}(-1)^{x+z}\langle A_x B_1 C_z\rangle\,$ $\textmd{and}\,\langle A_x B_y C_z\rangle$=$\sum\limits_{a, b, c}(-1)^{a+b+c}P(a, b, c|x,y, z).$ Here $A_x$, $B_y$ and $C_z$ are the observables for binary inputs $x,\,y,\,z$ of Alice, Bob and Charlie respectively. $a,b,c\in\{0,1\}$ denote the corresponding outputs.
\subsection{Tripartite Nonlocality}
Let $\rho_{ABC}$ denote a given tripartite quantum state shared between Alice, Bob and Charlie. The correlations arising due to measurements of the parties on their respective subsystems are local if those can be decomposed in the form:
\begin{equation}\label{pst2}
P(a,b,c|x,y,z)=\sum_{\lambda}p_{\lambda}P(a|x, \lambda)P(b|y,\lambda)]P(c|z,\lambda).
\end{equation}
The correlations which lack such a decomposition(Eq.(\ref{pst2})) are said to be nonlocal and the tripartite state $\rho_{ABC}$ is said to be nonlocal. The nonlocal correlations are capable of  violating  tripartite Bell inequality.
For such a tripartite scenario where each of the three parties can perform any one of two binary inputs,  a Bell-local polytope having $46$ facets was defined in \cite{Sliwa}. The set of $46$ inequalities, each defining a facet of the local polytope, serves as a necessary and sufficient criterion to detect nonlocality of tripartite correlations generated by local measurements by each of Alice, Bob and Charlie on their respective subsystems. So if $\rho_{ABC}$ fails to violate all $46$ facets then $\rho_{ABC}$ is local(upto this scenario\cite{Sliwa}).\\
We now deal with a weaker notion of steering nonlocality for a tripartite system\cite{caval1}
\subsection{Tripartite Steering Nonlocality}
As before, let $\rho_{ABC}$ be a state shared between Alice, Bob and Charlie. Let there be a referee who wants to check whether the correlations shared between the three parties are steerable. However he trusts the measurement apparatus of only one party, say Bob but does not trust that of Alice and Charlie. He will be convinced about the steerability of the tripartite correlations(from Alice and Charlie to Bob) if those correlations are inexplicable in the form:
\begin{equation}\label{steer1}
P(a,b,c|x,y,z)=\sum_{\lambda} q_{\lambda}P(a|x,\lambda)\textmd{Tr}[\widehat{\Pi}(b|y)\rho_{B}(\lambda)]P(c|z,\lambda).
\end{equation}
 Here $\lambda$ denotes the hidden variable, $x,z$ and $a,c$ denote local inputs and outputs of Alice and Charlie respectively. $\widehat{\Pi}(b|y)$ denotes the projection operator corresponding to observable
characterized  by  Bob's setting $y$ such that this is associated with the eigenvalue $b$. $\rho_{B}(\lambda)$, parameterized by the hidden variable $\lambda$,  stands for some pure state of Bob's system. If the tripartite correlations are steerable, i.e., cannot be decomposed in the above form(Eq.(\ref{steer1})), then the corresponding tripartite state $\rho_{ABC}$ is considered to be \textit{steerable} from Alice and Charlie to Bob. In \cite{caval1}, a Bell-type inequality (Eq.22 in \cite{caval1}) under some specific measurement settings(von-Neumann equatorial measurements) of all the parties(with referee trusting only one party) was given to detect steerability of $\rho_{ABC}$. Violation of this inequality is only sufficient to detect steerability of $\rho_{ABC}$ from two to one party(Bob and Charlie to Alice). So no definite conclusion can be given if it is satisfied.\\
Now we discuss a few details regarding detection of tripartite entanglement.
\subsection{Separability Criteria}
In \cite{guhne}, different criteria based on the density matrix formalism of  a tripartite state were given. Some of those criteria were necessary for a state to be separable. We discuss those criteria here.
 Let $(\rho_{l,m})_{8\times8}$ denote the eight dimensional density matrix corresponding to $\rho_{ABC}$. If the state is separable, then the elements $\rho_{l,m}$ necessarily satisfy the following criteria:
\begin{equation}\label{sep1}
    |\rho_{1,8}|\leq \sqrt[6]{\Pi_{i=2}^7\rho_{i,i}}
\end{equation}
\begin{equation}\label{sep2}
    |\rho_{1,8}|\leq \sqrt[6]{\rho_{1,1}*\rho_{4,4}*\Pi_{i=4}^7\rho_{i,i}}.
\end{equation}
So if at least one of these inequalities(Eqs.(\ref{sep1},\ref{sep2})) is violated then the tripartite state is entangled whereas if it satisfies both of these inequalities, then nothing can be said about its entanglement. \\
Having discussed mathematical prerequisites, we now present our discussion.
\section{Trilocal network scenario}\label{tri}
It is a network of five parties say Alice, Bob, Charlie, Dick and Tom(see FIG.2). They share three independent sources $S_1$, $S_2$ and $S_3$, characterized by the hidden variables $\lambda_1, ~ \lambda_2$ and $\lambda_3$ with independent probability distributions $\rho_1(\lambda_1), ~ \rho_2(\lambda_2)$ and $\rho_3(\lambda_3)$ respectively. Hence they satisfy:
 \begin{equation}\label{n6}
    \rho(\lambda_1,\lambda_2,\lambda_3)=\rho_1(\lambda_1)\rho_2(\lambda_2)\rho_3(\lambda_3)
\end{equation}
along with $\int d\lambda_i \rho_i(\lambda_i)=1(i=1,2,3)$.
Let $x, y, z, w, u (\in\{0,1\})$ denote the binary inputs and $a, b, c, d, e (\in\{0,1\})$ denote the outputs of Alice, Bob, Charlie, Dick and Tom respectively. Each party performs measurement on its respective particles. Each of Bob and Charlie receives three particles, one from each source($S_i$). Bob and Charlie are named as \textit{intermediate parties}. Each of the remaining three parties Alice, Dick and Tom receives only one particle and are termed as \textit{extreme} parties.  No communication is allowed between the parties. The correlations are \textit{trilocal} if they satisfy:
\begin{widetext}
\begin{equation}\label{n7}
P(a,b,c,d,e|x,y,z,w,u) = \iiint d\lambda_{1} d\lambda_{2}d\lambda_{3} {\rho(\lambda_1,\lambda_2,\lambda_3)}P(a|x, \lambda_1)P(b|y, \lambda_1, \lambda_2,\lambda_3)P(c|z, \lambda_1, \lambda_2,\lambda_3)P(d|w, \lambda_2)P(e|u, \lambda_3)
\end{equation}
\end{widetext}
together with the constraint given by Eq.(\ref{n6}). Under source independence restriction(Eq.(\ref{n6})), correlations which cannot be decomposed as above(Eq.(\ref{n7})) are said to be \textit{nontrilocal} in nature. By construction, set of trilocal correlations forms a subset of local correlations. However, the set of trilocal correlations is not convex due to the non-linear constraint(Eq.(\ref{n6})). Now if nontrilocal correlations are generated in the network when each of the three sources generates identical copy of a tripartite system say $S$, then $S$ is said to be $nontrilocal$ and this form of nonlocality is said to be \textit{nontrilocality} of $S$. To detect nontrilocal correlations we frame a set of sufficient criteria in the form of non-linear Bell-type inequalities.
\begin{figure}[htb]
\includegraphics[width=3.4in]{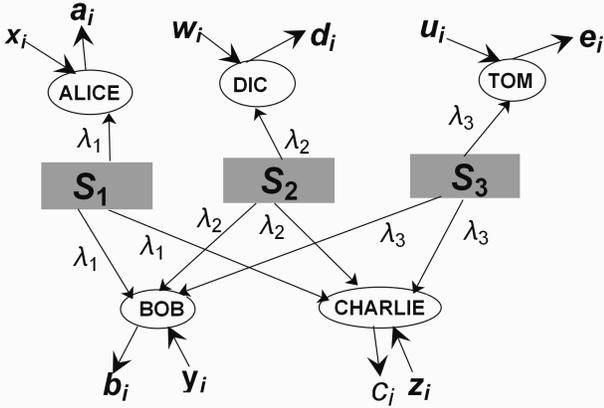}\\
\caption{\emph{Schematic diagram of trilocal network scenario.}}
\end{figure}
\\
\textit{Theorem.1}: Any trilocal five partite correlation necessarily satisfies:
\begin{equation}\label{n8t}
    \sqrt[3]{|I_{i_1,i_2,0}|}+\sqrt[3]{|I_{j_1,j_2,1}|}\leq1\,\,\forall\,i_1,\,i_2,\,j_1,\,j_2\,\in\{0,1\}\,\,\textmd{where,}
\end{equation}
\begin{equation}\label{n9t}
I_{i_1(j_1),i_2(j_2),k}=\frac{1}{8}\sum \limits_{x, w,u=0,1}(-1)^{k*l}\langle A_x B_{i_1(j_1)} C_{i_2(j_2)} D_w T_u\rangle,
\end{equation}
$ \langle A_x B_yC_z D_w T_u\rangle$=$\sum\limits_{a, b, c,d,e}(-1)^{m}P(a,b, c,d,e|x, y,z,w,u),~ k\in\{0,1\}$ where $l=x+w+u$ and $m=a+b+c+d+e$. $A_x$ denote the observables for inputs $x$ of Alice. $B_y,~ C_z, ~ D_w,~T_u$ are similarly defined. Each of these $16$ inequalities is tight in the sense that there exist correlations reaching the bound $1$(See Appendix.B). In this context, one should note that the notion of tightness discussed here is different(to be specific weaker) compared to the standard notion of tightness in respect of Bell inequalities where usually tightness of a Bell inequality refers to a specific feature that the inequality is a facet of a local polytope(hence defined by local correlations). Now all of these inequalities are only necessary but not sufficient criteria of trilocality, i.e., there may exist nontrilocal correlations that satisfy all of these inequalities. However, violation of at least one of these inequalities guarantees nontrilocality of the correlations. Hence violation of Eq.(\ref{n8t}) for at least one possible $(i_1,i_2,j_1,j_2)$ is sufficient to detect nontrilocality of the  correlations.\\
\textit{Proof:} See Appendix.A.\\
 Next we put forward a set of Bell-type inequalities which act as sufficient criteria to capture nonlocality of the correlations generated in this scenario.\\
\textit{Theorem.2}: Any local five partite correlation necessarily satisfies:
\begin{equation}\label{n8}
  |I_{i_1,i_2,0}|+|I_{j_1,j_2,1}|\leq1\,\,\forall\,i_1,\,i_2,\,j_1,\,j_2\,\in\{0,1\}.
\end{equation}
Just as in Theorem.1, here also each of $16$ inequalities is tight in the sense that there exist correlations for which equality is obtained in each of the inequalities. This set of inequalities being only a set of necessary but not sufficient criteria for locality, no proper conclusion can be drawn about the nature of correlations satisfying the whole set of inequalities. However violation of at least one of these inequalities from the set guarantees that the corresponding correlations generated in the network are nonlocal in nature. A comparison of the form of trilocal and local inequalities for every possible combination $(i_1,i_2,j_1,j_2)$ clearly indicates the inclusion of set of trilocal correlations inside the set of local correlations(see FIG.3) as was already interpreted before from the decomposition of trilocal and local correlations. Throughout our discussion we denote the set of $16$ trilocal inequalities(Eq.(\ref{n8t})) as $\mathcal{T}$ and the set of $16$ local inequalities(Eq.(\ref{n8})) as $\mathcal{L}$. \\
After sketching the trilocal network scenario along with inequalities capturing nature of five partite correlations terms $P(a,b,c,d,e|x,y,z,w,u)$ characterizing the network, we now explore quantum network scenario compatible with our scenario.\\
\begin{figure}[htb]
\begin{tabular}{|c|}
\hline
\includegraphics[width=2.2in]{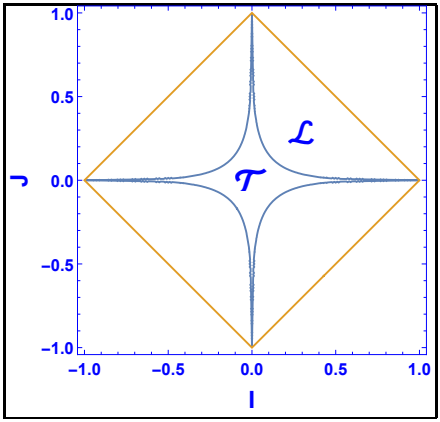}\\
\hline
\end{tabular}
\caption{\emph{Projection of the  correlation space in $(I,J)$ plane where $I=I_{i_1(j_1),i_2(j_2),0}$ and $J=I_{i_1(j_1),i_2(j_2),1}$ $\forall\,i_1,j_1,i_2,j_2\in\{0,1\}.$ $\mathcal{T}$ and $\mathcal{L} $ denotes the set of trilocal and local correlations respectively. Clearly $\mathcal{T}$ forms a proper subset of $\mathcal{L}$.}}
\end{figure}
\section{Quantum network}\label{qu}
Consider a double entanglement swapping network with three independent sources $S_1$, $S_2$ and $S_3$ each producing a tripartite entangled state. $S_1$ sends a tripartite state $\rho_{ABC}$ to Alice, Bob and Charlie. $S_2$ sends another tripartite state $\rho_{BCD}$ to Bob, Charlie and Dick. $S_3$ sends tripartite state $\rho_{BCT}$ to Bob, Charlie and Tom(see FIG.2). The overall quantum state is
\begin{equation}\label{n13}
\rho_{ABCDT}=\rho_{ABC}\small{\bigotimes}\rho_{BCD}\small{\bigotimes}\rho_{BCT}.
\end{equation}
The correlations arising due to measurements on $\rho_{ABCDT}$ characterize the network.
Each of the two intermediate parties, Bob and Charlie performs partial GHZ basis measurements on the joint state of three systems that each of them receives from the sources. For instance Bob performs partial GHZ basis measurement(see Appendix.C) on his three qubits(that he receives from $S_1$, $S_2$ and $S_3$), i.e., corresponding to each of the two inputs, he can distinguish between any two non overlapping groups formed from eight states of the GHZ basis. Similarly Charlie also performs partial GHZ basis measurements on the joint state of the three particles that he receives from the sources. Our choice for partial GHZ basis measurement against full GHZ basis is not only compatible with the nontrilocal network scenario(see FIG.3), but also justified from experimental perspectives as the latter is impossible in non-idealistic situations. Each of the three extreme parties, Alice, Dick and Tom  performs arbitrary projective measurements on their respective particles. The five partite correlators resulting due to measurements by the five parties on their respective subsystems are nontrilocal if those can violate at least one of $16$ inequalities(Eq(\ref{n8t})) from the set $\mathcal{T}$ whereas those are nonlocal if they can violate at least one of the inequalities(Eq.(\ref{n8})) from the set($\mathcal{L}$). If nontrilocal correlations are obtained when all three sources  $S_i(i=1,2,3)$ generate identical copy of a tripartite quantum state $\rho$(say) then that guarantees the state $\rho$ is nontrilocal. This form of nonlocality of a tripartite state is referred to as its $\textit{nontrilocality}$. Analogously if nonlocal correlations are generated in the network due to use of three  identical copies of $\rho$ then $\rho$ is said to be nonlocal in the network scenario(apart from standard Bell scenario).\\
We now proceed to discuss  quantum violation of inequalities from $\mathcal{T}$. Let $\mathfrak{B}^\mathcal{T}_{\rho}$ denote the upper bound of violation of at least one of the trilocal inequalities(Eq.(\ref{n8t})) from $\mathcal{T}$ by the correlations generated in the network where each of the three sources $S_1$, $S_2$ and $S_3$  produce identical copies of a tripartite quantum state $\rho.$ For our purpose we have used $\textit{Mathematica}$ software\cite{mathe} to find $\mathfrak{B}^\mathcal{T}_{\rho}$ for any $\rho.$. However, speaking of set of local inequalities given by Eq.(\ref{n8}), after thorough numerical observations we conjecture that quantum violation of local inequalities(Eq.(\ref{n8})) is impossible.
\subsection{Nontrilocal correlations from pure entanglement}
Let each of the three sources generates identical copy of pure tripartite state belonging to generalized Greenberger-Horne-Zeilinger(GGHZ) class of states\cite{ACN}:
\begin{equation}\label{S15}
 |\varphi_{GGHZ}\rangle\,=\,\cos(\alpha)|000\rangle+\sin(\alpha)|111\rangle,\,\alpha\in[0,\frac{\pi}{4}],
\end{equation}
Under suitable joint measurements by the intermediate parties(see Appendix.C) and projective measurements by the extreme parties, the upper bound of violation of Eq.(\ref{n8t}) is given by:
\begin{equation}\label{g1}
    \mathfrak{B}^\mathcal{T}_{\textmd{GGHZ}}=\sqrt[3]{2}\sin(2\alpha).
\end{equation}
Hence when each of the three sources produces any state belonging to GGHZ family characterized by the condition $\sin(2\alpha)>\frac{1}{\sqrt[3]{2}}$, nontrilocal correlations are generated in the network. Consequently all those states are nonlocal in the sense of revealing nontrilocality(as discussed before). Now the quantity $\sin^2(2\alpha)$ is the measure of genuine tripartite entanglement given by $3-$tangle $\tau$\cite{COF} of the corresponding GGHZ state. So all those states from the GGHZ family having $\tau>\frac{1}{\sqrt[3]{4}}$ are nontrilocal(see FIG. 4(i)).\\
However for generation of nontrilocal correlations in a network the sources need not produce a  resource as strong as  genuine entanglement. If each source generates biseparable quantum state then also nontrilocal quantum correlations can be generated under suitable choice of measurements. For instance, consider that each of $S_1$, $S_2$ and $S_3$  generates a  biseparable state of the form:
\begin{equation}\label{n14}
    |\Psi\rangle\,=\, (\cos(\eta)|00\rangle+\sin(\eta)|11\rangle)\small{\bigotimes} (\varsigma_1|0\rangle+\varsigma_2|1\rangle)
  \end{equation}
with $\eta\in[0,\frac{\pi}{4}]$ and $|\varsigma_1|^2+|\varsigma_2|^2=1$. Maximal violation of Eq.(\ref{n8t}) is given by:
\begin{equation}\label{b1}
    \mathfrak{B}^\mathcal{T}_{|\Psi\rangle}= \textmd{Max}[2^{\frac{4}{3}}|\varsigma_1\varsigma_2|\sin(2\eta),\sin (2\eta)\sqrt[3]{2|1-6(\varsigma_1\varsigma_2)^2|}].
\end{equation}
$ \mathfrak{B}^\mathcal{T}_{|\Psi\rangle}$ justifies our claim that violation of trilocal inequalities can also be achieved by using a weaker resource of biseparable entanglement(see FIG.4(ii)). \\
Having observed violation of trilocal inequalities by tripartite pure entangled states, we now explore whether nontrilocal correlations can be obtained from tripartite mixed entanglement.
\begin{center}
\begin{figure}
\begin{tabular}{|c|c|}
\hline
\subfloat[$|\varphi_{GGHZ}\rangle$]{\includegraphics[trim = 0mm 0mm 0mm 0mm,clip,scale=0.35]{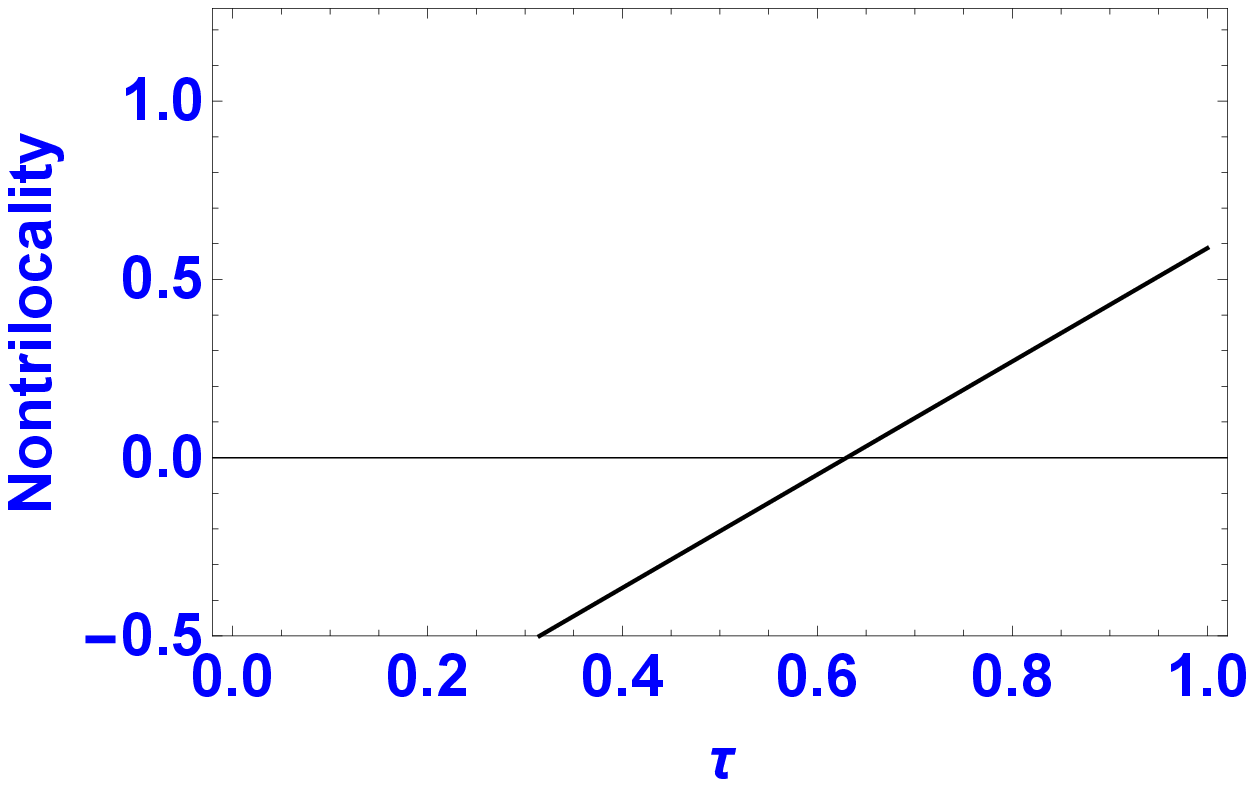}} &
\subfloat[$|\Psi\rangle$]{\includegraphics[trim = 0mm 0mm 0mm 0mm,clip,scale=0.35]{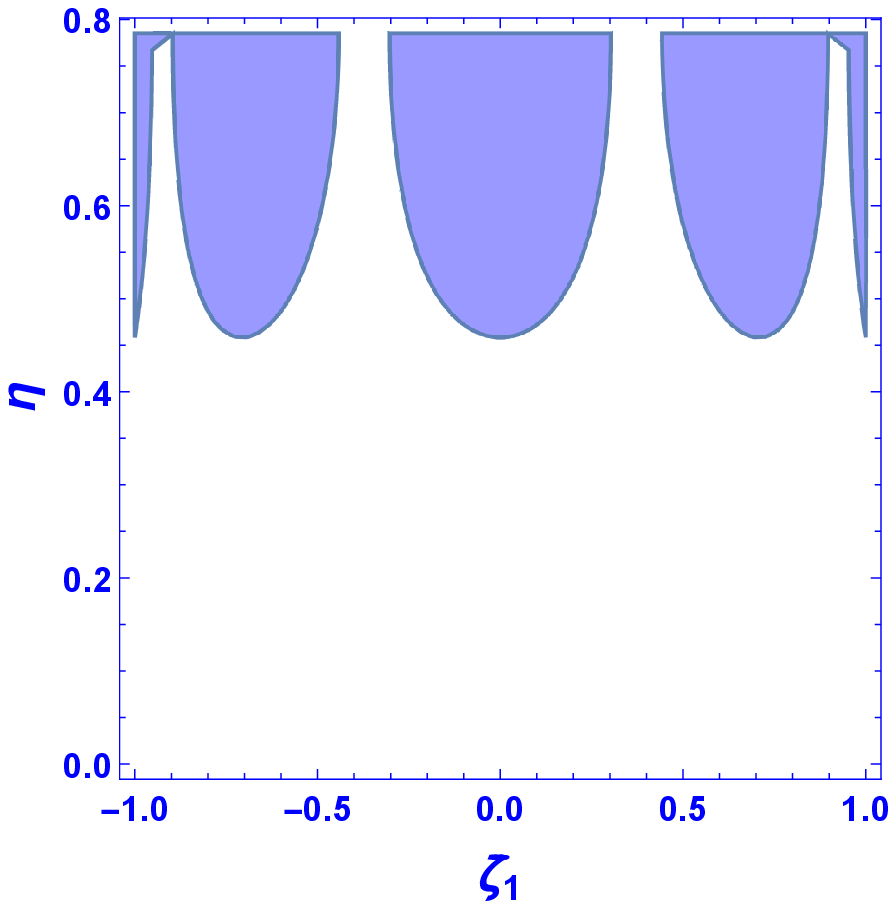}}\\
\hline
\end{tabular}
\caption{\emph{These two subfigures show quantum violation of trilocal inequalities(Eq.(\ref{n8t})) by pure tripartite entangled states. Figure (i) shows violation when each of the sources $S_1$, $ S_2$ and $S_3$ generates genuine entanglement whereas violation of Eq.(\ref{n8t}) by  biseparable entanglement is given in Figure (ii). }}
\end{figure}
\end{center}
\subsection{Nontrilocal correlations from mixed entanglement}
Consider the GHZ symmetric class of states, a family of mixed entangled states which has been extensively used in many recent research  topics\cite{gs1,gs2,gs3,gs4}. The family of states is given by:\\
$\Omega(p_1, p_2) = (\frac{2p_2}{\sqrt{3}}+ p_1)|GHZ_{+}\rangle\langle GHZ_{+}| \,+$
\begin{equation}\label{pst9iii}
(\frac{2p_2}{\sqrt{3}}- p_1)|GHZ_{-}\rangle\langle GHZ_{-}| + (1- \frac{4p_2}{\sqrt{3}})\frac{\mathbb{I}_8}{8}
\end{equation}
where
\begin{equation}\label{ghz}
    |GHZ_{\pm}\rangle = \frac{|000\rangle \pm |111\rangle}{\sqrt{2}}.
\end{equation}
and $\mathbb{I}_{8}$ denotes $8\times 8$ order identity matrix. The requirement $\Omega(p_1, p_2)\geq 0$ gives the constraints:
\begin{equation}\label{g1}
    -\frac{1}{4\sqrt{3}}\leq p_2 \leq \frac{\sqrt{3}}{4}
\end{equation}
 and
\begin{equation}\label{pst9iv}
|p_1| \leq \frac{1}{8}+\frac{\sqrt{3}}{2} p_2.
\end{equation}
Not only GHZ states but this family of states also include $\frac{\mathbb{I}_8}{8}$(maximally mixed state). The upper bound of violation of trilocal inequalities by this family is given by:
\begin{equation}\label{ghzs1}
    \mathfrak{B}^\mathcal{T}_{\Omega}=\sqrt[3]{16}|p_1|.
\end{equation}
So any mixed entangled state belonging to GHZ symmetric family having state parameter $p_1$ restricted by the condition $|p_1|>\frac{1}{\sqrt[3]{16}}$ can be used in the network to generate nontrilocal correlations(see FIG.5)). \\
However for each of these three families of tripartite entangled states(Eqs.(\ref{S15},\ref{n14},\ref{pst9iii})) giving quantum violation of trilocal inequalities, none of them can violate any of the local inequalities from the set $\mathcal{L}$. Hence the local inequalities(Eq.(\ref{n8})) may not be helpful for exploiting nonlocality in a quantum network. However, as already discussed before, the local inequalities(Eq.(\ref{n8})) being only necessary for a correlation to be local, if a correlation satisfies all $16$ inequalities given by Eq.(\ref{n8}) then no definite conclusion can be given. In this context, it may be mentioned that definite conclusion can be given if one can design a LHV(local hidden variable) model for such correlations. However, in absence of any such model and confining our discussion only upto these necessary criteria(Eq.(\ref{n8t}) for trilocality and Eq.(\ref{n8}) for locality) trilocal inequalities(Eq.(\ref{n8t})) seem to act as better detector of nonlocality(in sense of nontrilocality). This in turn gives a glimpse of the advantage of source independence(trilocal) assumption for revealing nonlocality in a quantum network in contrast to standard notions of nonlocality. At this junction we are now going  to discuss our findings related to the utility of trilocal assumption to exploit quantumness in a network compared to some usual procedures of doing so.\\
\begin{center}
\begin{figure}
\includegraphics[width=2.2in]{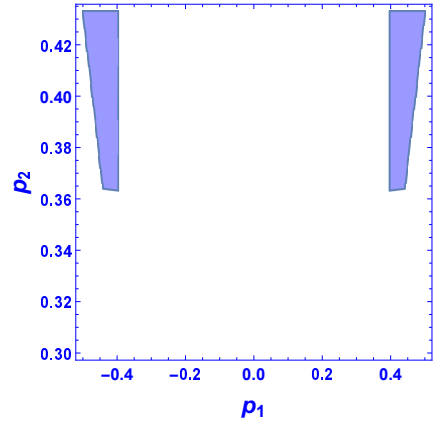}
\caption{\emph{The figure depicts a subspace in the parameter space $(p_1,p_2)$ of GHZ symmetric class of states(Eq.(\ref{pst9iii})). Shaded region gives the restrictions over state parameters $p_1$ and $p_2$ for which the corresponding states belonging to this family are nontrilocal, i.e, when used in the trilocal network, generate nontrilocal correlations. }}
\end{figure}
\end{center}
\section{Resistance To Noise}\label{res}
As we have pointed out in the last section, here we will present our observations related to trilocal assumption. Earlier works on source independence assumption\cite{BRA,BRAN} suggest that the most important physical interpretation of nontrilocal correlations is obtained by quantifying their resistance to noise.\\
Suppose each of the three sources produces a noisy state($\rho_{\textmd{noisy}}^i$). If $\nu_i$ denotes the visibility, a measure of resistance to noise offered by $\rho_{\textmd{noisy}}^i$ generated from $S_i(i=1,2,3)$, then the nature of quantum correlations in the network depends on $\mathcal{V}=\nu_1\nu_2\nu_3.$ Let \textit{trilocality threshold}($\Delta_{TRILOC}$) be defined as the largest value of $\mathcal{V}$ upto which the correlations do not violate Eq.(\ref{n8t}) for any possible $(i_1,i_2,j_1,j_2)$. Analogously we define the largest possible value of $\mathcal{V}$ upto which the correlations do not violate any of the local inequalities Eq.(\ref{n8}) from the set $\mathcal{L}$ as \textit{locality threshold}($\Delta_{LOC}$). In this context, it may also be mentioned that nonlocality in a quantum network can also be exploited by the correlations generated from the swapped states. To be specific, it may be checked whether depending on the output of partial GHZ basis measurement by the intermediate parties the conditional tripartite entangled state shared between the extreme parties generate nonlocal correlations. Such an exploration of nonlocality based on conditional tripartite entangled states is justified on the basis of observations obtained in a bipartite scenario in \cite{pircond}. In our scenario, after GHZ basis measurement performed by Bob and Charlie, a tripartite state(say $\chi$) is shared between the extreme parties Alice, Dick and Tom. Let $\mathcal{V}^C$ be defined as the  measure of resistance to noise offered by $\chi$ and let $\delta_{LOC}$ be interpreted as the largest possible value of $\mathcal{V}^C$ upto which $\chi$ is local in the specific Bell sense\cite{Sliwa}. Now if $\chi$ fails to violate all facets of the local polytope\cite{Sliwa} then that guarantees the state to be local in the corresponding Bell scenario. Intuitively, such a state is therefore unable to produce any genuine correlations and hence cannot be genuinely steerable from one party to other two parties\cite{jeva}. However it may be capable of generating only the weaker form of steering correlations\cite{caval1}. For this purpose, we analogously define $\delta_{NS}$ as the largest possible value of $\mathcal{V}^C$ upto which correlations obtained from $\chi$ may have a $VVS$ model where $V$ stands for local hidden variable and $S$ denotes local hidden state\cite{caval1}. So other than using any inequality from the set $\mathcal{L}$, i.e. any local inequality(Eq.(\ref{n8})), exploring nonlocality of the states resulting from  partial GHZ basis measurements by the intermediate parties and ultimately shared between the extreme parties can also capture nonlocality of correlations(if any) characterizing  corresponding network. These means of exploiting nonlocality and hence quantumness in a network by observing nonlocality of conditional states($\chi$) resulting in the network are however assisted with one way communication(from Bob and Charlie to extreme parties). Interestingly, we observed that even then there  exist instances where source independence assumption emerges to be more efficient compared to these notions of nonlocality. Below we present our findings in this context.
\subsection{Advantage of trilocal assumption}
In order to exploit the advantage of source independence assumption in a quantum network we consider the family of noisy GHZ states:
\begin{equation}\label{n16}
\vartheta=\epsilon  |GHZ_+\rangle\langle GHZ_+|+(1-\epsilon)\frac{\mathbb{I}_8}{8},\,\textmd{where}\,\epsilon\in[0,1]
\end{equation}
and $|GHZ_+\rangle$ is given by Eq.(\ref{ghz}). Here $\epsilon$ denotes the visibility of $\vartheta$.
This family is obtained when GHZ state($|GHZ_+\rangle$) is passed through a depolarization channel\cite{NIE,BEN,BNN,BRS}. Let each of the three sources $S_i$ generates a noisy state(Eq.(\ref{n16})) having visibility $\epsilon_i(i=1,2,3)$ respectively. After some suitable partial GHZ basis measurements by each of the two intermediate parties and suitable projective measurements by the extreme parties, the correlations generated in the  network are nontrilocal if:
 \begin{equation}\label{mix1}
    \epsilon_1\epsilon_2\epsilon_3>\frac{1}{2}.
 \end{equation}
 In this case, $V_{TRILOC}= \epsilon_1\epsilon_2\epsilon_3$. Clearly Eq.(\ref{mix1}) implies that $\Delta_{TRILOC}=\frac{1}{2}$, i.e. nonlocality(nontrilocality) is revealed for any value of  $V_{TRILOC}\in(\frac{1}{2},1]$. However none of the local inequalities can be violated by the correlations generated in this case. So definitely the set $\mathcal{T}$ can be considered more efficient compared to the set $\mathcal{L}.$ Now let the outputs of partial GHZ basis measurements by Bob and Charlie are communicated to the extreme parties. Here $\mathcal{V}^C=\epsilon_1\epsilon_2\epsilon_3$. It is observed that none of  $16$ swapped states $\chi_i^{DEP}$(say), can violate\cite{info} not even one of $46$ facets of the local polytope\cite{Sliwa} for any  value of $\mathcal{V}^C\leq 0.804$. Hence $\delta_{LOC}=0.804$. Clearly $\delta_{LOC}>\Delta_{TRILOC}$. Again none of $\chi_i^{DEP}(i=1,...,16)$ can violate Eq.(22) of \cite{caval1} and hence may not be steerable. So upto the existing detector of steering nonlocality(Eq.(22) of \cite{caval1}), $\delta_{NS}=1.$ Hence $\delta_{NS}>\Delta_{TRILOC}.$ This in turn justifies our claim that trilocal assumption emerges as a more efficient detector of nonlocality in a quantum network in contrast to these standard detectors of nonlocality.\\
 Till now we have discussed about the advantage of trilocal assumption in an entanglement swapping network, i.e., the network where the final states shared between the extreme parties are entangled. However this assumption can exploit quantumness(via nontrilocal correlations) even if the particles(qubits) forming the tripartite state(shared between the extreme parties) are atmost ppt bound entangled states\cite{review}. For instance, let each of $S_i$ produces identical copy of the noisy version of $|GHZ_+\rangle$ after it is passed through an amplitude damping channel\cite{NIE,MARK1}, say $\rho_{AMP}$(see Appendix.D). Let $\chi_k^{AMP}(k=1,...,16)$ denote the conditional tripartite states shared between the extreme parties depending on the outputs of the intermediate parties. Each of them satisfies the criteria necessary for separability(Eqs.(\ref{sep1},\ref{sep2})) and hence cannot be guaranteed to be entangled. Moreover negativity vanishes in all three cuts for each of these states($\chi_k^{AMP}(k=1,...,16)$). Hence each of them is either separable or atmost ppt bound entangled state. Also each of them is local\cite{Sliwa} and also produces correlations which may not be steerable, i.e. satisfy Eq.(22) of \cite{caval1}. But if the sources $S_i(i=1,2,3)$ are independent, then nontrilocal correlations are generated if:
 \begin{equation}\label{amp}
    \iota^3>\frac{1}{\sqrt[3]{4}},\,\,\textmd{where}
 \end{equation}
 $\iota=1-\gamma_A$($\gamma_A$ is the parameter characterizing the amplitude damping channel). Here $V_{TRILOC}=\iota^3$. Hence $\Delta_{TRILOC}=\frac{1}{\sqrt[3]{4}}$. Analogous observations are obtained if each of $S_i$ now produces identical copy of $\rho_{PHASE}$, where $\rho_{PHASE}$(see Appendix.D) denotes the noisy $|GHZ_+\rangle$ after being passed through phase damping channel\cite{NIE,MARK1}. Here nontrilocality is obtained if
  \begin{equation}\label{phase}
    \omega^3>\frac{1}{\sqrt[3]{4}},\,\, \textmd{with}
  \end{equation}
  $\omega=1-\gamma_P$($\gamma_P$ parameterizes the phase damping channel). Here $V_{TRILOC}=\omega^3$.
  Hence here also $\Delta_{TRILOC}=\frac{1}{\sqrt[3]{4}}$. So these observations point out the fact that irrespective of the entanglement content of the states resulting from partial GHZ basis measurements, nontrilocality and hence nonlocality(apart from standard Bell sense) is revealed in a quantum network only if the sources($S_1,S_2,S_3$) are assumed to be independent. Having introduced and thereby discussing related issues of trilocal network scenario involving three sources, we now generalize the scenario involving $n$ independent sources.
\section{$n$-local network scenario}\label{network}
In order to study role of multipartite($n$$\geq 4$) entanglement in a source independent network scenario in which the number of non interacting particles is also increased from $3$ to $n(\geq 4)$, the trilocal network scenario can be extended to $n$-local($n\geq 4$) network scenario. Such an extension requires increase in the number of independent sources from $3$ to $n$, each generating an $n$ partite state, together with increase in the number of parties from $5$ to $2n-1$. For $n=3$, this scenario corresponds to the trilocal network scenario. Some of these $2n-1$ parties receive only one particle whereas some receive more than one particle. Keeping analogy with trilocal network scenario, let us denote the former and latter type as \textit{extreme} and \textit{intermediate} parties respectively. Out of $2n-1$ number of parties, the number of intermediate parties is $n-1$ and that of extreme parties is $n.$  Let the intermediate parties be marked as $B_i(i=1,2,...,n-1)$ and the extreme parties be denoted as  $A_i(i=1,2,...,n)$ sharing $n$ independent sources $S_i(i=1,2,...,n)$ characterized by $\lambda_1,\,\lambda_2,\,...,\lambda_n$ respectively(see FIG.6). Each of $n$ sources, say $S_i$ sends a particle(characterized by $\lambda_i$) to each of $n-1$ intermediate parties and a particle to the extreme party $A_i$. So each of the intermediate parties $B_i$ receives $n$ particles whereas each of the extreme parties $A_i$ receives only one particle. Let $x_i,y_i\in\{0,1\}$ denote the binary input of $A_i(i=1,2,...,n)$ and $B_i(i=1,2,...,n-1)$ respectively. Let $a_i,b_i\in\{0,1\}$ denote the output of $A_i(i=1,2,...,n)$ and $B_i(i=1,2,...,n-1)$ respectively. The parties are not allowed to communicate. The $2n-1$ partite correlational terms are $n$-local if they satisfy:
\begin{widetext}
\begin{equation}\label{n20}
P(a_1,...,a_{n},b_1,...,b_{n-1}|x_1,...,x_{n},y_1,...,y_{n-1}) = \iint...\int d\lambda_{1}...d\lambda_n {\rho(\lambda_1,...,\lambda_n)}\Pi_{i=1}^nP(a_i|x_i, \lambda_i)\Pi_{k=1}^{n-1}P(b_k|y_k, \lambda_1,..., \lambda_n)
\end{equation}
together with the constraint:
\begin{equation}\label{nn}
    \rho(\lambda_1,\lambda_2,...,\lambda_n)=\Pi_{i=1}^n\rho_i(\lambda_i)
\end{equation}
where $\int d\lambda_i \rho_i(\lambda_i)=1\,\forall\,i\in\{1,2,...,n\}. $
 Otherwise the correlations are non $n$-local. The $n$-local inequalities are given by the following theorem.\\
 \textit{Theorem.3}: Any $n$-local $2n-1$ partite correlation term necessarily satisfies:
\begin{equation}\label{n21}
\sqrt[n]{|I_{i_1,...,i_{n-1},0}|}+\sqrt[n]{|I_{j_1,...,j_{n-1},1}|}\leq1\,\,\forall\,i_1,...,\,i_{n-1},\,j_1,...,\,j_{n-1}\,\in\{0,1\}\,\,\textmd{where},
\end{equation}
\begin{equation}\label{n22}
I_{i_1(j_1),...,i_{n-1}(j_{n-1}),k}=\frac{1}{2^n}\sum \limits_{x_1,..., x_n=0,1}(-1)^{k(x_1+...+x_{n})}\langle A_{x_1},..., A_{x_{n}}B_{y_{i_1}(y_{j_1})},...,B_{y_{i_{n-1}}(y_{j_{n-1}})}\rangle,\,k\in\{0,1\}
\end{equation}
and\\
$\langle A_{x_1},.., A_{x_{n}}B_{y_{i_1}(y_{j_1})},..,B_{y_{i_{n-1}}(y_{j_{n-1}})}\rangle= \sum\limits_{a_1,..,a_{n},b_1,..,b_{n-1}=0,1}(-1)^{g}P(a_1,..,a_n,b_1,..,b_{n-1}|x_1, ..,x_{n},y_{i_1}(y_{j_1}),..,y_{i_{n-1}}(y_{j_{n-1}}))$
\end{widetext}
where $g$$=$$a_1+...+a_{n}+b_1+...+b_{n-1}$. Here $A_{x_i}$ denotes the observable for binary inputs $x_i(i=1,...,n)$ of party $A_i$. $B_{y_{i_l}}$, $B_{y_{j_l}}(i_l,j_l=1,...,n-1)$ are similarly defined.
Violation of Eq.(\ref{n21}) for at least one possible $(i_1,i_2,...,i_{n-1},j_1,j_2,...,j_{n-1})$ guarantees non $n$-local nature of the correlations generated in the network. However the set $\mathcal{T}_n$ of $4^{n-1}$ $n$-local inequalities, being a set of necessary criteria for $n$-locality only, there may exist non $n$-local correlations satisfying all of $4^{n-1}$ inequalities. Similarly the local inequalities associated to $n$-local scenario are of the form:
\begin{equation}\label{n21l}
|I_{i_1,...,i_{n-1},0}|+|I_{j_1,...,j_{n-1},1}|\leq1\,\,\forall\,i_1,...,\,i_{n-1},\,j_1,...,\,j_{n-1}\,\small{\in}\{0,1\}.
\end{equation}
\\
 \begin{center}
\begin{figure}
\includegraphics[width=3.5in]{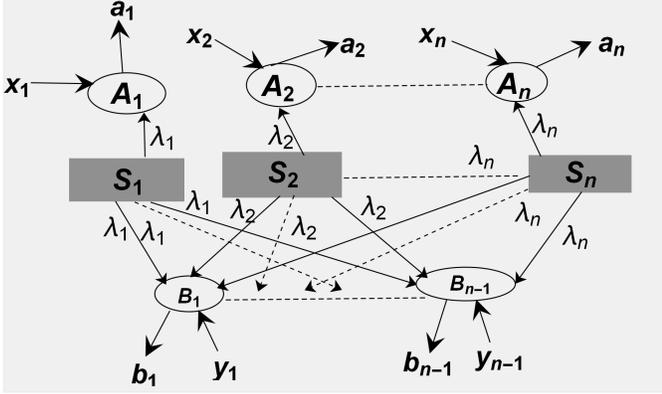}
\caption{\emph{Schematic diagram of a $n$-local network scenario. For $n=3$, this corresponds to the trilocal network scenario. }}
\end{figure}
\end{center}

In quantum scenario, consider a multiple($n-1$) entanglement swapping  network of $2n-1(n\geq4)$ parties sharing $n$ independent sources $S_i(i=1,...,n)$ such that each source produces an $n$-qubit state. Each of $n-1$ intermediate parties $B_1,...,B_{n-1}$ performs partial $n$ dimensional GHZ basis measurement on $n$ qubits($i^{th}$ qubit received from $S_i$). As a result of joint measurements by each of $n-1$ intermediate parties, an $n$ partite state is now shared between the extreme $A_i(i=1,...,n)$. Each of $A_i(i=1,...,n)$ now preforms projective measurement on its qubit. If non $n$-local correlations are generated in a network when all the sources generate identical copies of an $n$ partite state, then the corresponding $n$ partite state is said to be $non\,n-local$. At this junction it is expected that $n$-local assumption will give advantage over standard procedures of testing multipartite nonlocality. For instance if each of $n$ independent sources $S_i$ generates noisy version of $n$ partite GHZ state :
\begin{equation}\label{n20}
\vartheta_n^i=\epsilon_i |GHZ_+^n\rangle\langle GHZ_+^n|+(1-\epsilon_i) \frac{\mathbb{I}_{2^n}}{2^n},
\end{equation}
\textmd{where} $\epsilon_i\in[0,1]\,\forall i\in\{1,2,...,n\}$ and $n$ partite GHZ state $|GHZ_+^n\rangle$ is given by:
\begin{equation}\label{ghzn}
    |GHZ_+^n\rangle=\frac{|0,0,...0\rangle+|1,1,...,1\rangle}{\sqrt{2}}.
\end{equation}
We conjecture that non $n$-local correlations will be generated if:
\begin{equation}\label{con}
    \Pi_{i=1}^n \epsilon_i>\frac{1}{2}.
\end{equation}
Intuitively, nonlocality can be detected neither by any of $4^{n-1}$ local inequalities(Eq.(\ref{n21l})) nor by any standard procedure( as discussed for trilocal scenario) such as  testing nonlocality of the $n$ partite state resulting from partial GHZ basis measurement in standard Bell scenario or any other weaker form of nonlocality such as steering nonlocality.
\subsection{Comparing $n$-local network scenario with the network scenario in \cite{den}}
In \cite{den} the authors have given some general recipes  to design Bell-type inequalities so as to capture nonlocal behavior of correlations generated in network scenarios involving independent sources. The basic motivation behind their work was to design inequalities for an extended network($\mathcal{N}^{'}$,see FIG.1 of \cite{den}) involving say $m+1$ parties starting from a network of $m$ parties. To be more specific, starting from an already existing network $\mathcal{N}$, and a corresponding Bell-type inequality, they provided an iterative technique to design new Bell-type inequalities for a more complex network $\mathcal{N}^{'}$ such that the new network  involves one additional source($\mathcal{S}_{n+1}$, say) together with one additional observer($\mathcal{A}^{m+1}$,say). Their extension and hence their procedure of designing inequalities is characterized by the fact that $\mathcal{N}$ is extended to $\mathcal{N}^{'}$ via only one party($\mathcal{A}^{m}$) involved in $\mathcal{N}$ together with the assumption that $\mathcal{A}^{m}$ has binary outputs and $\mathcal{A}^{m+1}$ has binary inputs and outputs. It is to this party $\mathcal{A}^{m}$ that the additional source $\mathcal{S}_{n+1}$ sends a particle. Hence only one party acts a connector between $\mathcal{N}^{'}$ and $\mathcal{N}$. In \cite{den}, Rosset \textit{e.t al.} analyzed their procedures for some pre existing network scenarios such as bilocal network\cite{BRA,BRAN}, $n$- linear chain network, star network\cite{Tav}.\\
Now the $n$-local scenario(FIG.6) introduced in this paper is incompatible with the network scenario introduced in \cite{den}. This is because of the fact that $n$-local($n >2$) network introduced here cannot be considered as an extension of an existing network such that there is a single connecting party. To be more explicit, consider $n=3.$ Referring to FIG.2, each of the two intermediate parties Bob and Charlie receives particles from each of the three sources $S_1$, $S_2$ and $S_3.$ So without loss of any generality, if $S_3$ be considered as an additional source and Tom as an additional party, then $S_3$ sends particles to both of Bob and Charlie of the existing network(involving sources $S_1$, $S_2$ and parties Alice, Bob, Charlie and Dick) and hence the number of connecting parties is two. This creates a difference between the network scenario introduced in \cite{den} and that discussed here. Hence the technique for designing Bell-type inequalites\cite{den} becomes inapplicable here. As has already been discussed before, the motivation behind designing our $n$-local($n>2$) network is to explain quantum correlations resulting from an entanglement swapping protocol entangling $n$ non interacting particles such that each of the $n$ independent sources in the protocol generates an $n$-partite entangled state. Keeping this motivation intact and following the direction introduced in \cite{den}, it will be interesting if one can develop an iterative technique to design Bell-type inequalities for an extended network scenario starting from an existing one.
\section{Discussions}\label{dis}
To the best of authors' knowledge, till date, all the research activities on source independence assumption in a network dealt with characterization of correlations in the network where either the independent sources sent bipartite states to the parties\cite{BRAN,Tav} or the number of non interacting particles that got entangled finally was limited upto two\cite{raf}. In our present topic of discussion we have intended to consider both of these features simultaneously. For that we have introduced $n$-local network scenario where each of $n$ sources generate an $n$-partite state independently and the number of non interacting particles is also $n(>2)$. We have given a detailed discussion of the case where $n=3$, i.e., a trilocal network scenario where each of three sources generates a tripartite state and the number of extreme parties, holding the non interacting particles is particularly three. Based on  our findings as discussed above, it can safely be concluded that trilocal assumption reduces the requirements to reveal nonlocality of correlations characterizing a quantum network thereby emerging as a better tool to exploit nonlocality compared to some standard methods of doing so. However we have been able to provide with only some sufficient criteria to detect non $n$-locality. Further better results are expected if some necessary criteria for detecting the same can be given. Besides, randomness being an intrinsic property of quantum states, exploration of randomness associated with nonlocal correlations in a quantum network having independent sources($n$-local network) can be a topic of future research. Apart from theoretical perspectives, our present topic of discussion, basically being related with correlations in a quantum network, can be subjected to practical demonstrations and also can be used in related  experimental works\cite{ex1,ex2}.\\
 \textit{Acknowledgement:}
The authors acknowledge useful suggestions from an anonymous referee regarding some previous works related to the present topic  which have helped us to motivate this paper. The author D. Sarkar  acknowledges SERB, DST  India and DSA-SAP for financial support. The authors also acknowledge  fruitful discussions with Dr. A.Sen and SRF S.Karmakar.

\section{Appendix}
\subsection{Proof of Theorems}\label{A}
First we prove Theorem.1.\\
\textit{Proof}: Without loss of any generality, let $i_1=i_2=j_1=j_2=0.$
\begin{equation}\label{S1}
I_{0,0,0}=\frac{1}{8}\sum \limits_{x, w, u=0,1}\langle A_x B_{0} C_{0} D_w T_u\rangle
\end{equation}
and
\begin{equation}\label{S2}
I_{0,0,1}=\frac{1}{8}\sum \limits_{x, w,u=0,1}(-1)^{(x+w+u)}\langle A_x B_{0} C_{0} D_w T_u\rangle
\end{equation}
By our assumption, the conditional probability terms $P(a,b,c,d,e|x,y,z,w,u)$ admit trilocal decomposition(Eq.(\ref{n7})). Let us define marginal expectation:
\begin{equation}\label{S3}
    \langle A_x\rangle_{\lambda_1}=\sum_{a=0,1}(-1)^a P(a|x,\lambda_1).
\end{equation}
Other terms such as $\langle B_y\rangle_{\lambda_1,\lambda_2,\lambda_3}$, $\langle C_z\rangle_{\lambda_1,\lambda_2,\lambda_3}$, $\langle D_w\rangle_{\lambda_2}$ and $\langle T_u\rangle_{\lambda_3}$ are defined similarly. Using these terms  and trilocal assumption(Eq.(\ref{n6})) we get,
\begin{widetext}
$$|I_{0,0,0}|=\frac{1}{8}|\int\int\int d\lambda_1 d\lambda_2 d\lambda_3 \rho_1(\lambda_1)\rho_2(\lambda_2)\rho_3(\lambda_3)(\langle A_0\rangle_{\lambda_1}+\langle A_1\rangle_{\lambda_1})\langle B_0\rangle_{\lambda_1,\lambda_2,\lambda_3}\langle C_0\rangle_{\lambda_1,\lambda_2,\lambda_3}(\langle D_0\rangle_{\lambda_2}+\langle D_1\rangle_{\lambda_2})(\langle T_0\rangle_{\lambda_3}+\langle T_1\rangle_{\lambda_3})|$$

\begin{equation}\label{S4}
\leq \frac{1}{8}\int\int\int d\lambda_1 d\lambda_2 d\lambda_3 \rho_1(\lambda_1)\rho_2(\lambda_2)\rho_3(\lambda_3)|(\langle A_0\rangle_{\lambda_1}+\langle A_1\rangle_{\lambda_1})||\langle B_0\rangle_{\lambda_1,\lambda_2,\lambda_3}||\langle C_0\rangle_{\lambda_1,\lambda_2,\lambda_3}||(\langle D_0\rangle_{\lambda_2}+\langle D_1\rangle_{\lambda_2})(\langle T_0\rangle_{\lambda_3}+\langle T_1\rangle_{\lambda_3})|
\end{equation}
Now $|\langle B_0\rangle_{\lambda_1,\lambda_2,\lambda_3}|,|\langle C_0\rangle_{\lambda_1,\lambda_2,\lambda_3}|\leq 1$. Using these, Eq.(\ref{S4}) becomes,
$$|I_{0,0,0}|\leq \frac{1}{8}\int\int\int d\lambda_1 d\lambda_2 d\lambda_3 \rho_1(\lambda_1)\rho_2(\lambda_2)\rho_3(\lambda_3)|\langle A_0\rangle_{\lambda_1}+\langle A_1\rangle_{\lambda_1}||\langle D_0\rangle_{\lambda_2}+\langle D_1\rangle_{\lambda_2}||\langle T_0\rangle_{\lambda_3}+\langle T_1\rangle_{\lambda_3}|$$
\begin{equation}\label{S5}
  =\frac{\int d\lambda_1  \rho_1(\lambda_1)|\langle A_0\rangle_{\lambda_1}+\langle A_1\rangle_{\lambda_1}|}{2}\frac{ \int d\lambda_2  \rho_2(\lambda_2)|\langle D_0\rangle_{\lambda_2}+\langle D_1\rangle_{\lambda_2}|}{2}\frac{ \int d\lambda_3  \rho_3(\lambda_3)|\langle T_0\rangle_{\lambda_3}+\langle T_1\rangle_{\lambda_3}|}{2}
\end{equation}
Similarly one can get,
\begin{equation}\label{S6}
|I_{0,0,1}|\leq \frac{\int d\lambda_1  \rho_1(\lambda_1)|(\langle A_0\rangle_{\lambda_1}-\langle A_1\rangle_{\lambda_1})|}{2}\frac{ |\int d\lambda_2  \rho_2(\lambda_2)|(\langle D_0\rangle_{\lambda_2}-\langle D_1\rangle_{\lambda_2})|}{2}\frac{ \int d\lambda_3  \rho_3(\lambda_3)|(\langle T_0\rangle_{\lambda_3}-\langle T_1\rangle_{\lambda_3})|}{2}
\end{equation}
Now for any $6$ positive integers, $m,m^{'},m^{''},s,s^{'},s^{''}$, the inequality $\sqrt[3]{mm^{'}m^{''}}+\sqrt[3]{ss^{'}s^{''}}\leq\sqrt[3]{m+s}\sqrt[3]{m^{'}+s^{'}}\sqrt[3]{m^{''}+s^{''}}$ holds(Holder's Inequality).
 Using this inequality on the upper bounds of $|I_{0,0,0}|$(Eq.\ref{S5}) and $|I_{0,0,1}|$(Eq.(\ref{S6})) we get,
$$\sqrt[3]{|I_{0,0,0}|}+\sqrt[3]{|I_{0,0,1}|}\leq \sqrt[3]{\int d\lambda_1  \rho_1(\lambda_1)\frac{|\langle A_0\rangle_{\lambda_1}+\langle A_1\rangle_{\lambda_1}|}{2}+\frac{|\langle A_0\rangle_{\lambda_1}-\langle A_1\rangle_{\lambda_1}|}{2}}$$
\begin{equation}\label{S7}
\times\sqrt[3]{\int d\lambda_2  \rho_2(\lambda_2)\frac{|\langle D_0\rangle_{\lambda_2}+\langle D_1\rangle_{\lambda_2}|}{2}+\frac{|\langle D_0\rangle_{\lambda_2}-\langle D_1\rangle_{\lambda_2}|}{2}}\sqrt[3]{\int d\lambda_3  \rho_3(\lambda_3)\frac{|\langle T_0\rangle_{\lambda_3}+\langle T_1\rangle_{\lambda_3}|}{2}+\frac{|\langle T_0\rangle_{\lambda_3}-\langle T_1\rangle_{\lambda_3}|}{2}}
\end{equation}
Now $\frac{|\langle A_0\rangle_{\lambda_1}+\langle A_1\rangle_{\lambda_1}|}{2}+\frac{|\langle A_0\rangle_{\lambda_1}-\langle A_1\rangle_{\lambda_1}| }{2}$ $ =$ $\small{\textmd{max}}\{|\langle A_0\rangle_{\lambda_1}|,|\langle A_1\rangle_{\lambda_1}|\}\leq 1.$ Similarly for observables $D_0$ and $D_1$,  $\frac{|\langle D_0\rangle_{\lambda_2}+\langle D_1\rangle_{\lambda_2}|}{2}+\frac{|\langle D_0\rangle_{\lambda_2}-\langle D_1\rangle_{\lambda_2}|}{2}$ $=$ $\small{\textmd{max}}\{|\langle D_0\rangle_{\lambda_2}|,|\langle D_1\rangle_{\lambda_2}|\}\leq 1$ and for observables $T_0$ and $T_1$,  $\frac{|\langle T_0\rangle_{\lambda_3}+\langle T_1\rangle_{\lambda_3}|}{2}+\frac{|\langle T_0\rangle_{\lambda_3}-\langle T_1\rangle_{\lambda_3}|}{2}$ $=$ $\small{\textmd{max}}\{|\langle T_0\rangle_{\lambda_3}|,|\langle T_1\rangle_{\lambda_3}|\}\leq 1$
\end{widetext}
Hence Eq.(\ref{S7}) becomes,
\begin{equation}\label{S8}
    \sqrt[3]{|I_{0,0,0}|}+\sqrt[3]{|I_{0,0,1}|}\leq 1.
\end{equation}
This proof holds for any possible combination of $(i_1,i_2,j_1,j_2).$ Hence trilocal correlations necessarily satisfy all of $16$ possible inequalities(Eq.(\ref{n8})). $\blacksquare$\\
\textit{Proof of Theorem.2}: This theorem can be proved in a similar pattern. The only difference is that instead of Holder's inequality, we have to use the following inequality involving positive numbers:
\begin{equation}\label{l1}
    mm^{'}m^{''}+ss^{'}s^{''}\leq (m+s)(m^{'}+s^{'})(m^{''}+s^{''}).
\end{equation}
\subsection{Tightness of the inequalities}\label{B}
Each of $16$ trilocal inequalities(Eq.(\ref{n8t})) is tight. To prove that we give an explicit trilocal decomposition of correlations for which equality holds in inequality given by Eq.(\ref{n8t}) for every possible combination $(i_1,i_2,j_1,j_2)$, i.e., for all $16$ inequalities. Let the correlation shared by the five parties be of the form:
\begin{center}
\begin{eqnarray*}
P(a|x,\lambda_1,\tau_1) &=& 1\,\, \textmd{if}\, a=\lambda_1\small{\bigoplus}\tau_1\ast x, \\
 &=& 0 \, \textmd{else}
\end{eqnarray*}
\begin{eqnarray*}
 P(d|w,\lambda_2,\tau_2) &=& 1\,\, \textmd{if}\, d=\lambda_2\small{\bigoplus}\tau_2\ast w, \\
 &=& 0 \, \textmd{else}
\end{eqnarray*}
\begin{eqnarray*}
 P(e|u,\lambda_3,\tau_3) &=& 1\,\, \textmd{if}\, e=\lambda_3\small{\bigoplus}\tau_3\ast u, \\
 &=& 0 \, \textmd{else}
\end{eqnarray*}
\begin{eqnarray*}
  P(b|y,\lambda_1,\lambda_2,\lambda_3) &=& 1\,\, \textmd{if}\, b=(\lambda_1*\lambda_3\small{\bigoplus}\lambda_2*\lambda_3)\\
   &=& 0 \, \textmd{else}
\end{eqnarray*}
\begin{eqnarray*}
  P(c|z,\lambda_1,\lambda_2,\lambda_3) &=& 1\,\, \textmd{if}\, c=\lambda_2\small{\bigoplus}\lambda_1*\lambda_3\\
   &=& 0 \, \textmd{else}
\end{eqnarray*}
\begin{eqnarray*}
  \rho_i(\lambda_i=0) &=& 1 \\
   &=& 0,\,\textmd{else}\,\,\,\forall i=1,2,3
\end{eqnarray*}
\begin{eqnarray*}
  \kappa_i(\tau_i=0) &=& r \\
  \kappa_i(\tau_i=1) &=& 1-r\,\,\,\forall i=1,2,3,\,r\in[0,1]
\end{eqnarray*}
\end{center}
 $\tau_1$, $\tau_2$ and $\tau_3$ are the sources of local randomness of Alice, Dick and Tom respectively with $r\in[0,1]$. For this form of correlations, $I_{i_1,i_2,0}=r^3$ $\forall i_1,i_2\in\{0,1\}$ and $I_{j_1,j_2,1}=(1-r)^3$ $\forall j_1,j_2\in\{0,1\}$. Hence $ \sqrt[3]{|I_{i_1,i_2,0}|}+\sqrt[3]{|I_{j_1,j_2,1}|}=1\,\,\forall\,i_1,\,i_2,\,j_1,\,j_2\,\in\{0,1\}.$\\
\subsection{Joint Measurements performed by the intermediate parties for quantum violation}
The GHZ basis is given by:
\begin{equation}\label{ap1}
|\phi_{mnk}\rangle=\frac{1}{\sqrt{2}}\sum_{l=0}^1(-1)^{m*l}|l\rangle|l\small{\bigoplus} n \rangle |l\small{\bigoplus} k\rangle,\,m,n,k\in\{0,1\}
\end{equation}
As stated in the main text each of the two intermediate parties(Bob and Charlie) performs partial GHZ basis measurements. To be specific  each of two binary valued measurements corresponds to distinguishing between two non overlapping groups of GHZ states. Let each of two measurements by Bob be denoted by $B_1$ and $B_2$. Let $G_{B_1}^1$ and $G_{B_1}^2$ denote two groups of GHZ basis elements corresponding to two outputs of the binary valued measurement $B_1$, i.e. Bob performing measurement $B_1$ means that he can distinguish between the two groups $G_{B_1}^1$ and $G_{B_1}^2$ of GHZ states($G_{B_1}^1$ vs $G_{B_1}^2$). $G_{B_2}^1$ and $G_{B_2}^2$ similarly denote two groups of GHZ elements corresponding to two outputs of the measurement input $B_2$. The measurement inputs and outputs for Charlie  are analogously defined: $G_{C_1}^1$ and $G_{C_1}^2$ denote outputs for input $C_1$ and $G_{C_2}^1$ and $G_{C_2}^2$ denote outputs for input $C_2$. In Table(I) we enlist the measurement settings for each of Bob and Charlie for which quantum violations are obtained when each of the sources generate a copy of GGHZ  states(Eq.(\ref{S15})).
\begin{table}[htp]
\begin{center}
\begin{tabular}{|c|c|}
\hline
$B_1$&$P_{000}+P_{001}+P_{010}+P_{100}-(P_{011}+P_{101}+P_{110}+P_{111})$\\
\hline
$B_2$&$P_{000}+P_{001}+P_{110}+P_{011}-(P_{010}+P_{100}+P_{101}+P_{111})$\\
\hline
$C_1$&$P_{000}+P_{001}+P_{010}+P_{100}-(P_{011}+P_{101}+P_{110}+P_{111})$\\
\hline
$C_2$&$P_{101}+P_{110}+P_{000}+P_{011}-(P_{010}+P_{100}+P_{001}+P_{111})$\\
\hline
\end{tabular}\\
\caption{The table gives the partial GHZ basis measurements by each of Bob and Charlie for GGHZ class of states(Eq.(\ref{S15})). Here $P_{ijk}$ denotes the projection operator along $|\phi_{ijk}\rangle$(Eq.(\ref{ap1})) $\forall i,j,k\in\{0,1\}$. As a result of their joint measurements and suitable projective measurements by Alice, Dick and Tom on their respective particles, the five partite correlations generated in the network are nontrilocal. }
\end{center}
\label{table1}
\end{table}

\subsection{Amplitude and Phase Damping Channels}
Here we discuss two noisy channels used in the main text.
\subsubsection{Amplitude Damping}
Amplitude damping channel's mathematical representation involves Krauss operators\cite{kra}. After a qubit state $\omega$ is passed through an amplitude damping channel, corresponding noisy version of the state is given by:
\begin{equation}\label{r2}
    \theta_{AMP}(\omega)=W_0\omega W_0\dag + W_1\omega W_1\dag
\end{equation}
 where $W_0=\left(
              \begin{array}{cc}
                1 & 0 \\
                0 & \sqrt{1-\gamma} \\
              \end{array}
            \right)$
and $W_1=\left(
              \begin{array}{cc}
                0 & \sqrt{\gamma} \\
                0 & 0 \\
              \end{array}
            \right)$
are the Krauss operators used for representing the operation of amplitude damping.  When each of the three qubits of a tripartite state($\varrho$) is passed individually through an amplitude damping channel characterized by parameter $\gamma_A$, noisy version of the state is given by:
\begin{equation}\label{krauss}
    \varrho_{AMP}=P( W_i\small{\bigotimes} W_j\small{\bigotimes} W_k)\varrho P( W_i\dag\small{\bigotimes} W_j\dag\small{\bigotimes} W_k\dag)
\end{equation}
where $P(W_i\small{\bigotimes} W_j\small{\bigotimes} W_k)$ denotes possible permutations of the operators $W_i,W_j,W_k$ over all possible $i,j,k\in\{0,1\}.$ When  $\varrho=|GHZ_+\rangle\langle GHZ_+|$(Eq.(\ref{ghz})), density matrix of the corresponding noisy state is given by:
 \begin{widetext}
\begin{equation}\label{pst12}
\rho_{AMP}=\left(\begin{array}{cccccccc}
\frac{1+\gamma_A^3}{2}&0&0&0&0&0 &0&\frac{\sqrt[3]{\iota^3}}{2}\\
0&\frac{\iota*\gamma_A^2}{2} & 0&0&0&0&0&0\\
0&0 & \frac{\iota*\gamma_A^2}{2}&0&0&0&0&0\\
0&0&0&\frac{\iota^2*\gamma_A}{2}&0&0&0&0\\
0&0&0&0&\frac{\iota*\gamma_A^2}{2}&0&0&0\\
0&0&0&0&0&\frac{\iota^2*\gamma_A}{2}&0&0\\
0 &0&0&0&0&0&\frac{\iota^2*\gamma_A}{2} &0\\
\frac{\sqrt[3]{\iota^3}}{2}&0&0&0&0&0&0&\frac{\iota^3}{2}\\
\end{array} \right)
\end{equation}
where $\iota=1-\gamma_A$ measures the resistance to noise by $\rho_{AMP}.$
\end{widetext}
\subsubsection{Phase Damping}
The Krauss operators involved in the formulation of a phase damping channel are:
  $Q_0=\left(
              \begin{array}{cc}
                1 & 0 \\
                0 & \sqrt{1-\gamma_P} \\
              \end{array}
            \right)$\\
and $Q_1=\left(
              \begin{array}{cc}
                0 & 0 \\
                0 & \sqrt{\gamma_P} \\
              \end{array}
            \right)$.\\
Analogous to amplitude damping channel, if each of the three qubits of $\varrho$ is passed individually through a phase damping channel characterized by parameter $\gamma_P$, noisy version of the state is given by Eq.(\ref{krauss}) only with the operators $W_0$ and $W_1$ being replaced by the operators $Q_0$ and $Q_1$.
So for  $\varrho=|GHZ_+\rangle\langle GHZ_+|$(Eq.(\ref{ghz})), corresponding noisy state( density matrix formalism) is represented as:
 \begin{widetext}
\begin{equation}\label{pst12}
\rho_{PHASE}=\left(\begin{array}{cccccccc}
\frac{1}{2}&0&0&0&0&0 &0&\frac{\sqrt[3]{\omega^3}}{2}\\
0&0 & 0&0&0&0&0&0\\
0&0 &0&0&0&0&0&0\\
0&0&0&0&0&0&0&0\\
0&0&0&0&0&0&0&0\\
0&0&0&0&0&0&0&0\\
0 &0&0&0&0&0&0&0\\
\frac{\sqrt[3]{\omega^3}}{2}&0&0&0&0&0&0&\frac{1}{2}\\
\end{array} \right)
\end{equation}
with $\omega=1-\gamma_P$ measuring the resistance to noise by $\rho_{PHASE}.$
\end{widetext}

\begin{thebibliography}{1}
\bibitem{info}For violation of each of $46$ inequalities, we have optimized over all projective measurements.
\end{thebibliography}
\end{document}